\documentclass{aa}  

\usepackage{graphicx}
\usepackage{hyperref}
\usepackage{txfonts}
\usepackage{xcolor}
\usepackage{caption}
\usepackage{subcaption}
\usepackage{textcomp}
\usepackage[rightcaption]{sidecap}
\usepackage[mathlines]{linenoaa}
\usepackage{placeins}  
\usepackage{media9}

\urldef{\meddeadata}\url{https://umbra.nascom.nasa.gov/padre/padre-meddea/l0/photon/}
\urldef{\demosunkitcustom}\url{https://sunkit-spex.readthedocs.io/en/latest/generated/gallery/legacy/fitting_custom_spectra.html}
\urldef{\meddeagsw}\url{https://github.com/PADRESat/padre_meddea/tree/main}

\begin{document} 

    \title{First results from PADRE/MeDDEA: X-rays from a flare and its CME}
    \titlerunning{First results from PADRE/MeDDEA}      
        
    \author{Muriel Zoë Stiefel\inst{1,2}\corrauth{muriel.stiefel@fhnw.ch}
            \and
            Säm Krucker\inst{1}\email{samuel.krucker@fhnw.ch}
            \and
            Olivier Limousin\inst{3}\email{Olivier.LIMOUSIN@cea.fr}
            \and
            Niharika Godbole\inst{4,5}\email{niharika.godbole@nasa.gov}
            \and
            Gwendoline Marc\inst{3}\email{gwendoline.marc@cea.fr}
            \and
            Amir Caspi\inst{6}\email{amir@boulder.swri.edu}
            \and
            Kyle Gregory\inst{4}\email{kyle.j.gregory@nasa.gov}
            \and
            Ace Stratton\inst{7}\email{ace.stratton@berkeley.edu}
            \and
            Diana Renaud\inst{3}\email{diana.renaud@cea.fr}
            \and
            Aline Meuris\inst{3}\email{aline.meuris@cea.fr}
            \and 
            Laura A. Hayes\inst{8}\email{laura.hayes@dias.ie}
            \and
            André Csillaghy\inst{1}\email{andre.csillaghy@fhnw.ch}
            \and
            Pascal Saint-Hilaire\inst{7}\email{shilaire@berkeley.edu}
            \and
            Juan Camilo Buitrago Casas\inst{7}\email{milo@ssl.berkeley.edu}
            \and 
            Christopher Smith\inst{7}\email{csmith@ssl.berkeley.edu}
            \and
            Savannah Sky Perez-Piel\inst{7}\email{savannah_sky@berkeley.edu}
            \and
            Anton Tremsin\inst{7}\email{astr@berkeley.edu}
            \and
            Elizabeth Warren\inst{7}\email{lizi.warren@berkeley.edu}
            \and
            Bradley Pafchek\inst{7}\email{bradley.pafchek@berkeley.edu}
            \and 
            Eliad Peretz\inst{4}\email{eliad.peretz@nasa.gov}
            \and 
            Carlos Urdiales\inst{9}\email{Carlos.Urdiales@swri.org}
            \and
            Reynaldo Gonzalez\inst{9}\email{Reynaldo.Gonzalez@swri.org}
            \and
            Chris M. Moeckel\inst{7}\email{chris.moeckel@berkeley.edu}
            \and
            Juan Carlos Martínez Oliveros\inst{7}\email{oliveros@ssl.berkeley.edu}
            \and
            Steven Christe\inst{4}\email{steven.d.christe@nasa.gov}
    }

    \institute{
            University of Applied Sciences and Arts Northwestern Switzerland, Bahnhofstrasse 6, 5210 Windisch, Switzerland
        \and    
            ETH Zürich, Rämistrasse 101, 8092 Zürich, Switzerland
        \and
            Université Paris-Saclay, Université Paris Cité, CEA, CNRS, AIM, 91191 Gif-sur-Yvette,France
        \and
            Heliophysics Science Division, NASA Goddard Space Flight Center, Greenbelt, MD 20771,USA
        \and
            American University, Washington, D.C., United States
        \and
            Southwest Research Institute, 1301 Walnut Street, Suite 400, Boulder, CO 80302, USA
        \and
            Space Sciences Laboratory, University of California, 7 Gauss Way, 94720 Berkeley, USA
        \and
            Astronomy \& Astrophysics Section, Dublin Institute for Advanced Studies, Dublin, D02 XF86, Ireland
        \and
            Southwest Research Institute, 6220 Culebra Road, San Antonio, TX 78238, USA
    }
        
    \date{Received July, 2026; accepted Month, year}
        
    \abstract 
    {The MeDDEA hard X-ray spectrometer onboard NASA's PADRE CubeSat operates four flight-spare detectors from the Solar Orbiter/STIX instrument.
    }
    {The aim of this work is to evaluate the accuracy of the MeDDEA calibration and to demonstrate that MeDDEA data are ready for scientific exploitation. Using the first joint observations between MeDDEA and STIX, we investigated faint HXR emission associated with an escaping CME relative to the emission from the full flare. 
    }  
    {In this work, the first large flare -- SOL2025-11-28T22, a GOES M6 class flare -- observed by MeDDEA was analyzed and the MeDDEA spectrum was compared with simultaneous observations from Fermi/GBM and ASO-S/HXI using the latest calibrations. From the Solar Orbiter viewpoint, the flare is highly limb-occulted, occurring 15$^\circ$ behind the solar limb and allowing STIX to isolate high coronal emission associated with the CME while MeDDEA observed the full flare. We combined spectral fitting with HXR/EUV imaging from STIX, HXI, and AIA to investigate the locations and energetics of the flare- and CME-associated HXR sources.
    }
    {The MeDDEA spectral observations are in agreement with FERMI/GBM and ASO-S/HXI within 10\% in the energy range of 18--30~keV. The CME-associated photon flux observed by STIX in the nonthermal range is faint at about 1\% of the flare photon flux. The energy content of the thermal component associated with the CME, on the other hand, is found to be of the same order of magnitude as the thermal energy of the flare loop, at least for a filling factor of unity. Imaging information from STIX in combination with AIA reveals that the emission measured by STIX is located within the associated CME. 
    }
    {We report that MeDDEA provides scientifically reliable spectral observations and is ready for scientific use. Joint observations with STIX using limb-occultation allowed us to measure the fraction of nonthermal electrons injected into the CME relative to the electrons in flares, demonstrating the scientific value of combining occulted and unocculted views of the same eruption and highlighting the scientific value of MeDDEA. 
    }

    \authorrunning{Stiefel, M. Z. et al.}
        \maketitle 
    \nolinenumbers

\section{Introduction}\label{sec:introduction}

Solar eruptions are driven by the release of stored magnetic energy in the solar corona. The release process is impulsive, producing not only heated plasma, but a significant fraction of the released energy goes into particle acceleration, forming high-energy tails in the particle distribution functions. This "flare" process forms flare ribbons in the chromosphere and hot, coronal flare loops that connect the flare ribbons \citep[for a review see][]{Benz_2017}. In addition, solar eruptive events eject plasma, energetic particles, and magnetic fields into the interplanetary space and drive a shock; this process is called coronal mass ejection \citep[CME; for a review, see][]{Webb_2012}. 

Hard X-ray (HXR) bremsstrahlung produced by nonthermal electrons is a powerful diagnostic of the acceleration process in solar eruptions, providing quantitative insights such as the energy content in accelerated electrons. Within solar eruptions, by far the strongest nonthermal HXR sources are from the flare ribbons in the chromosphere, where bremsstrahlung is strongest due to the much higher ambient densities compared to the density in the corona \citep[for a review see][]{Fletcher_2011}. Fainter, nonthermal bremsstrahlung sources are also omnipresent in the corona \citep[e.g.,][]{Krucker_Lin_2008}, but their detection is much more difficult due to the lower ambient density in the corona producing much less intense bremsstrahlung signals \citep[for a review see][]{Krucker_2008}. Furthermore, the presence of bremsstrahlung from thermal flare loops often outshines nonthermal bremsstrahlung from accelerated electrons in the corona. Electrons injected into the escaping CME produce an even fainter bremsstrahlung signal as the ambient density is even lower and the rapid expansion of the CME might decrease the density further. Generally, the flare HXR sources are so much brighter that the HXR emission associated with the CME is lost in the dynamic range of current spectral and imaging telescopes, which use indirect imaging systems, i.e., Fourier imagers. HXR emissions from the CME can only be detected in partially limb-occulted flares where the solar limb blocks HXR emission from the chromosphere and lower corona \citep[see][for a list of well-observed events]{Lastufka_2019}. However, for limb-occulted observations, the properties of the HXRs from the flare ribbons remain undetected, making it challenging to compare the HXR emissions or relative fluxes of the flare and CME together. 

With the Spectrometer/Telescope for Imaging X-rays \citep[STIX;][]{Krucker_2020} onboard Solar Orbiter \citep[][]{Muller_2020}, we now have, for the first time, a solar-dedicated HXR telescope making observations away from the Earth-Sun line, allowing us to combine occulted observations with an Earth-based HXR telescope that detects the entire flare. \citet{Krucker_2026} presented a single event study from two different vantage points, combining occulted STIX observations with Earth-orbiting HXR observatories, revealing that, despite their faint nature, HXR emissions from CMEs are an integral part of solar eruptions.

The solar PolArization and Directivity X-Ray Experiment (PADRE; Martinez et al. sub. to A\&A 2026) CubeSat mission was designed to investigate the angular distribution of the high-energy tails of flare-accelerated electrons, whether these electrons are beamed along the magnetic-field lines or whether they are isotropic, thus forming a cloud. A beamed electron distribution may produce a polarized primary signal with a characteristic directivity dependence that can be measured as a directivity signal \citep[e.g.][]{Jeffrey_2024}, while an isotropic distribution does not produce polarization and radiates in all directions with the same signal strength. PADRE provides measurements of the polarization and directivity using two independent HXR measurements. The Solar HARd X-ray Polarimeter instrument (SHARP; Buitrago Casas et al. in prep.) measures the degree of polarization of the HXR signal, while the Measuring Directivity to Determine Electron Anisotropy instrument (MeDDEA; Christe et al. sub. to A\&A 2026) measures the total HXR flux from the Sun. In combination with Solar Orbiter's different vantage points, MeDDEA and STIX investigate the directivity of the X-ray signal. MeDDEA and STIX fly the same CdTe detectors \citep[SO-Caliste;][]{Meuris_2012,Limousin_2016}, simplifying the relative detector calibration significantly. 
With these two independent measurements, the PADRE mission should provide a definite answer on the degree of beaming, depending on the availability of joint observations of a large flare with significant nonthermal emission and an appropriate viewing geometry. 

Besides the main science objective of measuring the degree of electron beaming in solar flares, the complementary vantage points of MeDDEA and STIX provide a more complete view of solar eruptions. For partially limb-occulted events, STIX can detect faint, high coronal, and CME-associated X-ray emission, while the much brighter emission from the main flare is blocked by the solar limb. At the same time, MeDDEA observes the event without occultation and measures the flare’s total X-ray emission. Combining these two views therefore allows the faint CME-associated component to be isolated and quantified relative to the full flare. The inverse scenario, observing an event occulted from Earth but visible to Solar Orbiter on-disk, is just as likely. While constraining directivity requires specific viewing angle geometries (Martinez et al. sub. to A\&A 2026, Christe et al. sub. to A\&A 2026), limb-occulted observations can occur at any time, except during those few days when Solar Orbiter is close to the Earth--Sun line, providing a rich dataset. Furthermore, the accuracy requirements of inter-instrument calibration knowledge are significantly reduced compared to those required to constrain electron anisotropy \citep[][]{Jeffrey_2024}. Hence, an occulted flare analysis with MeDDEA and STIX provides a path to highlight the science capabilities of MeDDEA.

This paper provides the first joint observations of a solar eruption captured by both the MeDDEA HXR spectrometer and the STIX HXR imaging spectrometer and provides constraints on the HXR signal from within the escaping CME compared to the full flare emission. The event presented here is the first large MeDDEA flare observed after the end of its commissioning phase, i.e., SOL2025-11-28T22 at a GOES M6 level, for which STIX observed the flare from an occultation angle of $\sim$15$^\circ$. In addition to MeDDEA, on-disk observations of the full flare are also available from other HXR instruments, namely the Hard X-ray Imager \citep[HXI;][]{Zhang_2019} on the Advanced Space-based Solar Observatory \citep[ASO-S;][]{Gan_2023}, the Gamma-ray Burst Monitor \citep[GBM;][]{Meegan_2009} on Fermi, Konus on WIND \citep[][]{Aptekar_1995}, and the High Energy L1 Orbiting X-ray Spectrometer \citep[HEL1OS;][]{Nandi_2025} on Aditya-L1 \citep[][]{Tripathi_2023}, allowing us to assess the total flux accuracy of MeDDEA relative to other observatories. Furthermore, ASO-S/HXI provides imaging information of the flare sources relative to the CME-associated sources imaged by STIX. EUV observations from the Atmospheric Imaging Assembly \citep[AIA;][]{Lemen2012aia} and white-light observations by the Helioseismic Magnetic Imager \citep[HMI;][]{Scherrer_2012,Schou2012}, both onboard the Solar Dynamics Observatory \citep[SDO;][]{pesnell2012sdo}, were used to set HXR imaging results into the context of the erupting event.

\section{Methods}

In the following section, we discuss the most important steps from the data analysis. We split the discussion into two main parts: the analysis of the data by the STIX, HXI, and GBM instruments; and a more detailed discussion of the analysis of the MeDDEA data. We leave out the spectral analysis of the HEL1OS and Konus data sets due to the large time-bin size and the limited time coverage, respectively. 

\subsection{Analysis of STIX, HXI, and GBM data}\label{Sec: Method data all instruments}

STIX, HXI, and GBM all observe the Sun in the high-energy range above 4 keV, including thermal and nonthermal emissions from solar flares. For the analysis of the STIX, HXI, and GBM data, we followed the same steps discussed in detail in \citet{Krucker_2026}. All times reported by STIX in this paper are corrected by the light-travel time from Solar Orbiter's position, which is 0.89AU to 1AU. The closer distance of Solar Orbiter is also taken into account in spectral fitting by scaling the flux to 1AU. 

The spectral fitting of all three instruments was done within the Object Spectral Executive \citep[OSPEX software;][]{Tolbert_2020} package available in IDL. This has been the standard spectral fitting tool for solar HXR spectra for nearly three decades, and it supports STIX, HXI, and GBM data products. For the STIX data, we applied the newest calibration, as presented in \citet[][]{Massa_2026}, available in \texttt{sswidl} version v0.6.2 as of July 2026.
For the image reconstructions with STIX and HXI, we used the standard \texttt{sswidl} routines for both instruments \citep[][]{Massa_2023, Su_2019}, and we used the maximum entropy method (\texttt{mem\_ge}) \citep[][]{Massa_2020} and CLEAN \citep[e.g.,][]{Catalano_2026} reconstruction methods. The images we show in this paper were reconstructed using CLEAN.

\subsection{Analysis of MeDDEA data}

MeDDEA photon event list data are made available online as soon as they are downlinked from the observatory on the NASA Solar Data Analysis Center website in FITS files\footnote{\meddeadata}. One file typically covers around 1~hour of observation time and records each photon count, containing its energy and timestamp. The MeDDEA team makes available all of their processing and analysis software in a Python package hosted on Github\footnote{\meddeagsw}.
In order to calibrate these data for scientific analysis, the data need to be energy calibrated,\footnote{Level 2 data will already be energy calibrated.} and the flux must be livetime corrected. Additionally, during large flares, the event list is sparsified on board in order to maintain high-energy events in the presence of high thermal fluxes. 
During the peak of the flare presented in this paper, MeDDEA went into level 2 decimation (this only affects photons with energies below $\sim$28~keV; see Christe et al. sub. to A\&A 2026 for details) and the relative livetime was above 0.75, a value for which the livetime correction works well. After correction, the photon event list can be binned in time and energy to create light curves and spectra. Errors are given by Poisson statistics.

Spectral analysis of MeDDEA was performed using \texttt{sunkit-spex} \citep[][]{ryan_2026}, the Python package for high-energy spectroscopy. Similarly to OSPEX, \texttt{sunkit-spex} uses forward-fitting of photon models to the observed count spectrum with a detector response matrix (DRM). Several studies have already shown that the results of \texttt{sunkit-spex} are reliable and consistent with the OSPEX fitting routines \citep[e.g.,][]{Bajnokova_2024, Stiefel_2026}.  
MeDDEA was in high-sensitivity mode for the flare presented in this paper, with all pixels using a low-energy threshold of $\sim$3~keV (see Christe et al. sub. to A\&A 2026). For simplicity, we only used the large pixels to derive the flare spectrum. To fit the MeDDEA spectrum in \texttt{sunkit-spex}, we followed the tutorial for fitting a customized spectrum\footnote{\demosunkitcustom}. A dedicated MeDDEA reader for the \texttt{sunkit-spex} environment is under development.

\section{Observations}\label{sec:observations}

   \begin{figure}
   \centering
   \includegraphics[width=9cm]{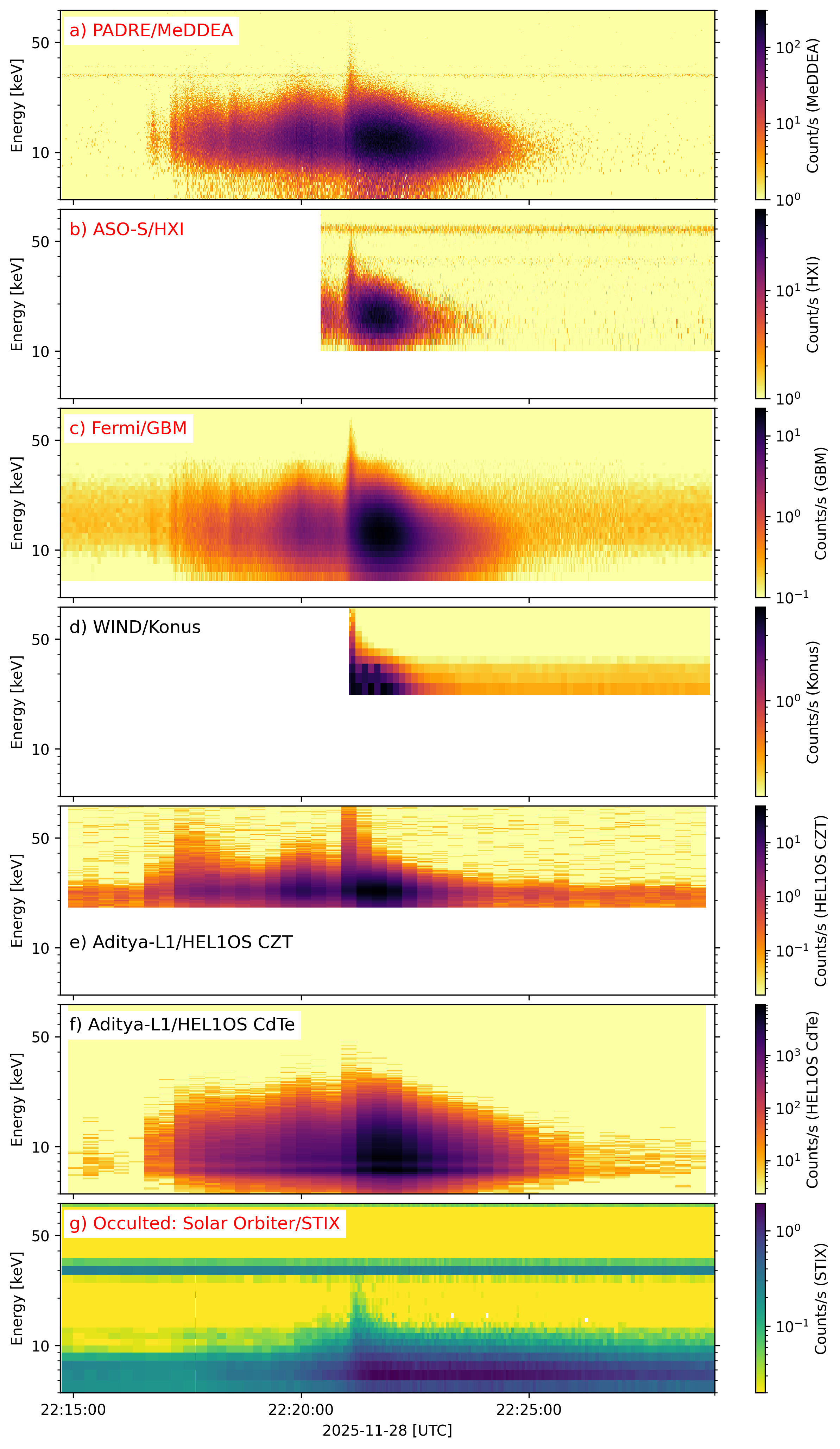}
   \caption{Spectrograms as a function of time of the M6-class flare on 28 November, 2025, observed by six different HXR instruments. From top to bottom: (a) PADRE/MeDDEA, (b) ASO-S/HXI, (c) Fermi/GBM, (d) WIND/Konus, (e) Aditya-L1/HEL1OS CZT- and (f) CdTe-detectors, and (g) Solar Orbiter/STIX spectrograms. The instruments labeled in red are used for further analysis in this paper. STIX sees the flare limb-occulted, all other instruments see the flare on disk. The spectrograms are not background subtracted; therefore, onboard calibration source lines are visible.
   }
   \label{Fig: Overview spectrograms}
    \end{figure}

In this paper, we present the M6-class flare on 28 November, 2025 at 22:20~UT, the first large flare recorded by MeDDEA. The flare was associated with a narrow CME at speeds of $\sim$700~km/s reported by the LASCO CME catalog and the CACTus CME catalog \citep[][]{Robbrecht_2004}. 
HMI difference images reveal that SOL2025-11-28 is a white-light flare with a source located above the limb at [-928.4{\arcsec},-293.1{\arcsec}], indicating that the HXR flare ribbons are visible from Earth \citep[for a detailed discussion of this type of flares see][]{Krucker_2015}.  
For Solar Orbiter, the flare was limb-occulted such that only the higher corona associated with the erupting CME was observed from its vantage point. As the flare position from Earth view is known, the occultation angle is also known ($\theta_{occ} = 15^\circ$), and the occultation height can be calculated at 24.5~Mm. For definitions of the occultation angle and height, see \citet[][]{Krucker_2026}, Eq. (1). In the following, we discuss the temporal evolution, spectral analysis, and imaging of the eruption, placing MeDDEA data into context with other instruments and comparing flare and CME HXR emissions.

\subsection{Temporal evolution}

Spectrogram plots of all HXR instruments that have observed the flare are shown in Fig.~\ref{Fig: Overview spectrograms}. The same energy range is shown for all instruments. The flux values are given in log scale using a single color table but with adapted scalings for the different instruments; this is with the exception of STIX, where a different color table is used to highlight that the signal is occulted. The MeDDEA spectrogram is shown in the top panel~(a), without instrumental background subtraction, making the calibration line visible around 30.85~keV ($^{133}$Ba). Panel~(b) shows the HXI spectrogram, also without background subtraction. The HXI calibration source, $^{241}$Am, is visible around 59.54~keV. Before the flare onset, HXI was in the South Atlantic Anomaly. These data are not shown for a clearer presentation, which is the reason for the white space. In panel~(c), the spectrogram of Fermi/GBM detector~N0 is shown. This detector has the lowest count rate of all GBM detectors (i.e., the Sun is at the edge of the detector field of view), and therefore pileup effects are of less concern compared to the other detectors that show at least a factor of~eight higher count rates. As the GBM sensitivity of a single detector is still high, we only used the GBM detector~N0 in our analysis to minimize pileup effects. Panel~(d) shows the Konus--Wind spectrogram from the triggered data product; therefore, no data are available before the trigger onset. Konus--Wind observes flares up to the gamma-ray range, but only the data up to 80~keV are shown here. Panels~(e) and~(f) show the spectrograms of HEL1OS on Aditya-L1. HEL1OS has two different detector sets: CZT, which is sensitive to high-energy ranges, i.e., mainly nonthermal emission; and CdTe, which is sensitive to lower energy ranges, i.e., mainly thermal emission. The plots shown use spectrogram data products available online, which are binned in predefined energy-bin widths and time bins of 20~s; higher resolution data are currently not publicly accessible. Panel~(g) shows the STIX spectrogram without background subtraction, revealing the $^{133}$Ba calibration line at 30.85~keV. The STIX times have been adjusted by +48.3~s to correct for the different light travel time due to the closer radial distance to the flare site. STIX does not observe the main flare on-disk, it only observes the occulted signal. The difference in the observations is discussed in the next paragraph. Figure~\ref{Fig: Overview spectrograms} demonstrates that MeDDEA provides high-quality data of solar flares at high temporal cadence and high energy resolution. 

   \begin{figure}
   \centering
   \includegraphics[width=9cm]{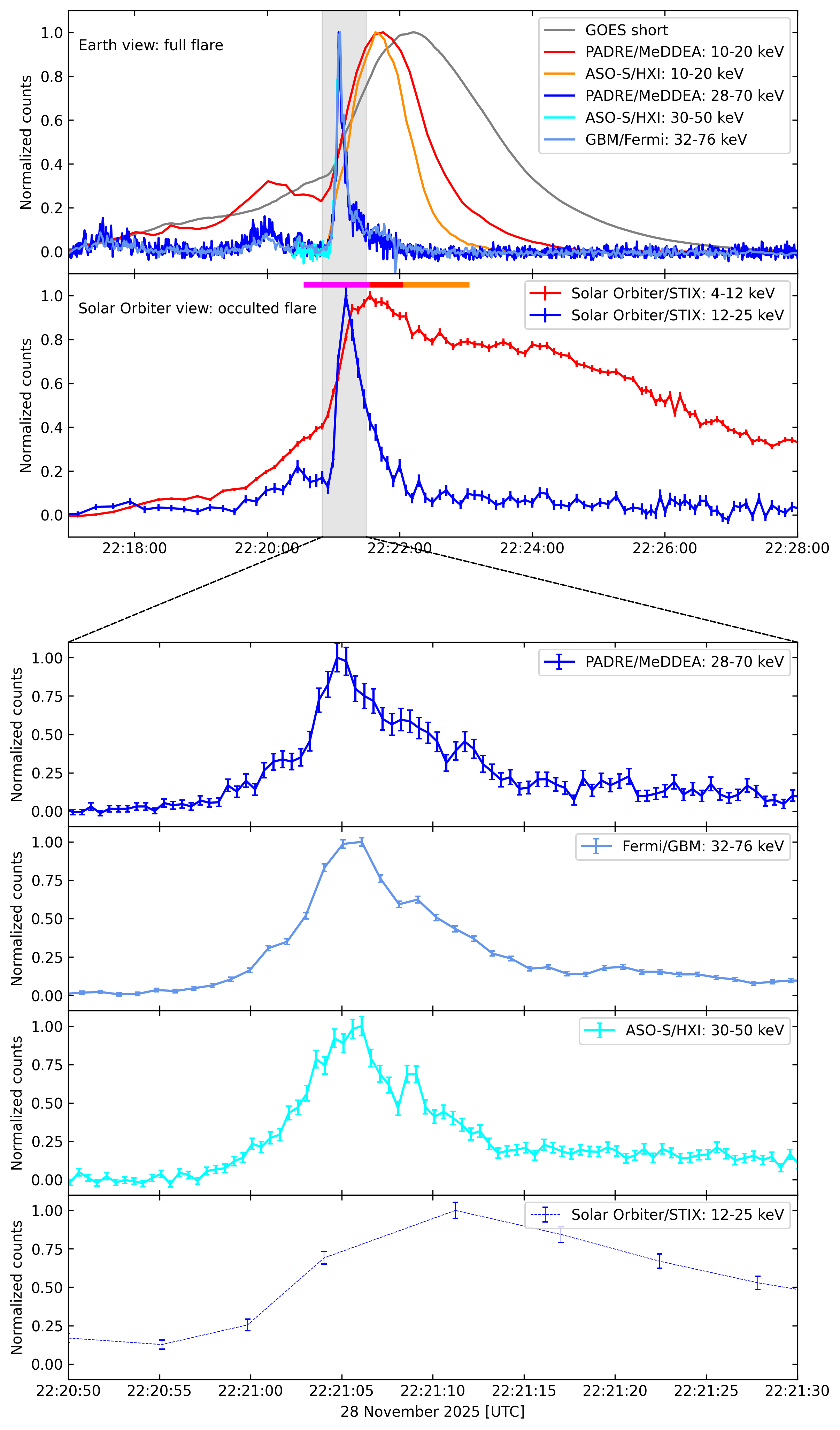}
   \caption{X-ray time profiles of the M6-class flare on 28 November, 2025 at different energies as indicated. In the top panel, light curves from instruments observing the full flare are shown. In the second panel, Solar Orbiter/STIX profiles are displayed showing the emission from the high corona. The three colored bars show the time ranges used for the image reconstructions discussed in Section~\ref{Sec: HXR imaging}. In the lower four plots, the nonthermal light curves of the four instruments are shown in more detail for the impulsive phase of the flare. The displayed time range is marked by the gray area in the top two panels.}
   \label{Fig: Overview lightcurves}
    \end{figure}

   \begin{figure*}
   \centering
   \includegraphics[width=18cm]{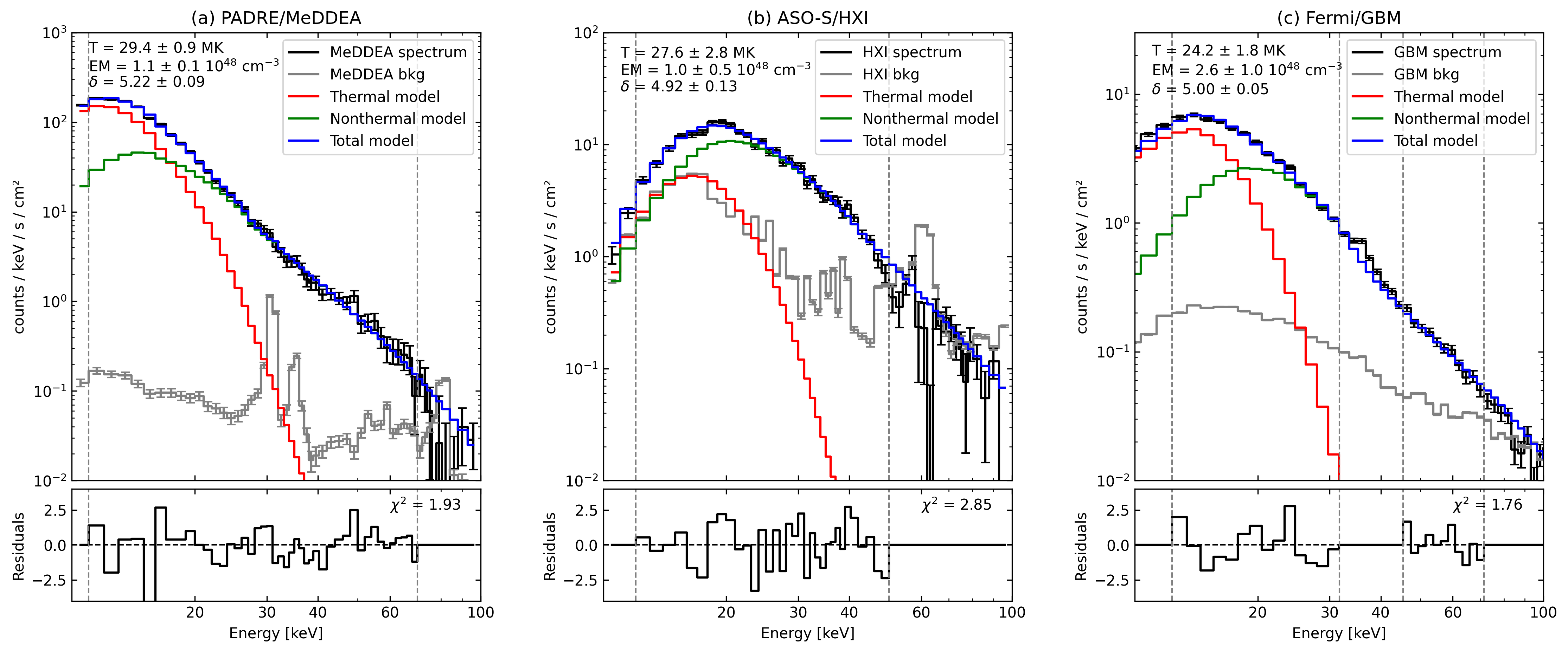}
   \caption{Observed count spectra around the nonthermal peak (22:21:02--22:21:10~UTC). From left to right: the spectrum of (a) PADRE/MeDDEA, (b) ASO-S/HXI, and (c) Fermi/GBM. In the top panels, the count spectra are shown in black with error bars. Each spectrum is fitted with a thermal (red) and a nonthermal (green) model. The sum of the two models, i.e., the total fit, is shown in blue. In the top left corner, the temperature ($T$) and the spectral index ($\delta$) of the thermal and nonthermal fits are shown. Below each spectrum, the residuals and the $\chi^2$ of the fit are reported. The dashed gray lines indicate the energy range used for fitting.}
   \label{Fig: Comparison spectra}
    \end{figure*}

Figure~\ref{Fig: Overview lightcurves} shows the extracted light curves of specific energy bands from MeDDEA and compares them to GOES/XRS, HXI, GBM, and STIX light curves. The lower panels show the nonthermal peak observed by the individual instruments in more detail. For this flare, time synchronization for MeDDEA was found to require a correction. A time shift of 5.0~s was required to match the peaks of the nonthermal emission between MeDDEA, HXI, and GBM. The selection for the displayed nonthermal energy ranges in Fig.~\ref{Fig: Overview lightcurves} is driven by excluding thermal emissions, but optimizing the count statistics for the nonthermal component for each instrument. As the spectral resolution for the individual instruments is different, this results in slightly different energy ranges for MeDDEA, HXI, and GBM. The nonthermal (28--70~keV) light curve of MeDDEA is shown with 0.5~s cadence, resulting in relative error bars at peak time of 9\%. In the nonthermal range, all three instruments show a similar time evolution within error bars, as expected. The thermal HXR profiles peak earlier than the GOES/XRS short wavelength profile, as generally observed. This indicates that the hottest plasma, which dominates the measurements by MeDDEA and HXI, peaks earlier during the course of a flare compared to the cooler plasma, which dominates the GOES measurements. Due to the thicker entrance window of HXI \citep[][]{Zhang_2019} compared to MeDDEA, the 10--20~keV count range corresponds to a higher average photon energy for HXI, and therefore the HXI 10--20~keV curve peaks and decays before the MeDDEA profile.

Due to the limb-occultation, the light curves observed by STIX look different compared to the other instruments that observed the entire event. The STIX nonthermal time profile at 12--25~keV has a similar onset, but it is less impulsive and has a longer decay time. This is typical for highly occulted events, which mainly see HXRs associated with CMEs \citep[see][]{Lastufka_2019}. For the strongest, highly occulted events, a clear exponential decay can be observed over minutes \citep[e.g.][]{Krucker_2007}, suggesting that electrons are trapped within the magnetic structure of the CME, and the exponential decay is indicative of the different energy-loss processes. In events with strong losses (e.g., collisional losses can dominate due to high ambient densities), the time evolution of the high coronal source becomes similar to that observed in the flare ribbons \citep{Krucker_2026}. For the event shown here, the decay is rather fast, but nevertheless longer than for the associated flare, suggesting that this event is in between the two extreme scenarios. The thermal profile of the occulted emission is clearly different from the flare profile. During the nonthermal burst, the thermal emission is increasing---reminiscent of the Neupert effect \citep[][]{Neupert_1968, Veronig_2002}---suggesting that nonthermal electrons heat the high coronal source, as reported by \citet{Glesener_2013} for a different event. While the emission of the presented flare decays rather quickly, the occulted signal is slowly decaying. The fast flare decay is likely driven by conductive losses, which are rather strong for the compact flare loop observed in this flare. Conductive losses for the high coronal source seen in the occulted signal, on the other hand, are much less important and therefore occur on longer timescales. In the next sections, we discuss the relative intensities of the occulted emission to the entire flare. 

   \begin{figure}
   \centering
   \includegraphics[width=9cm]{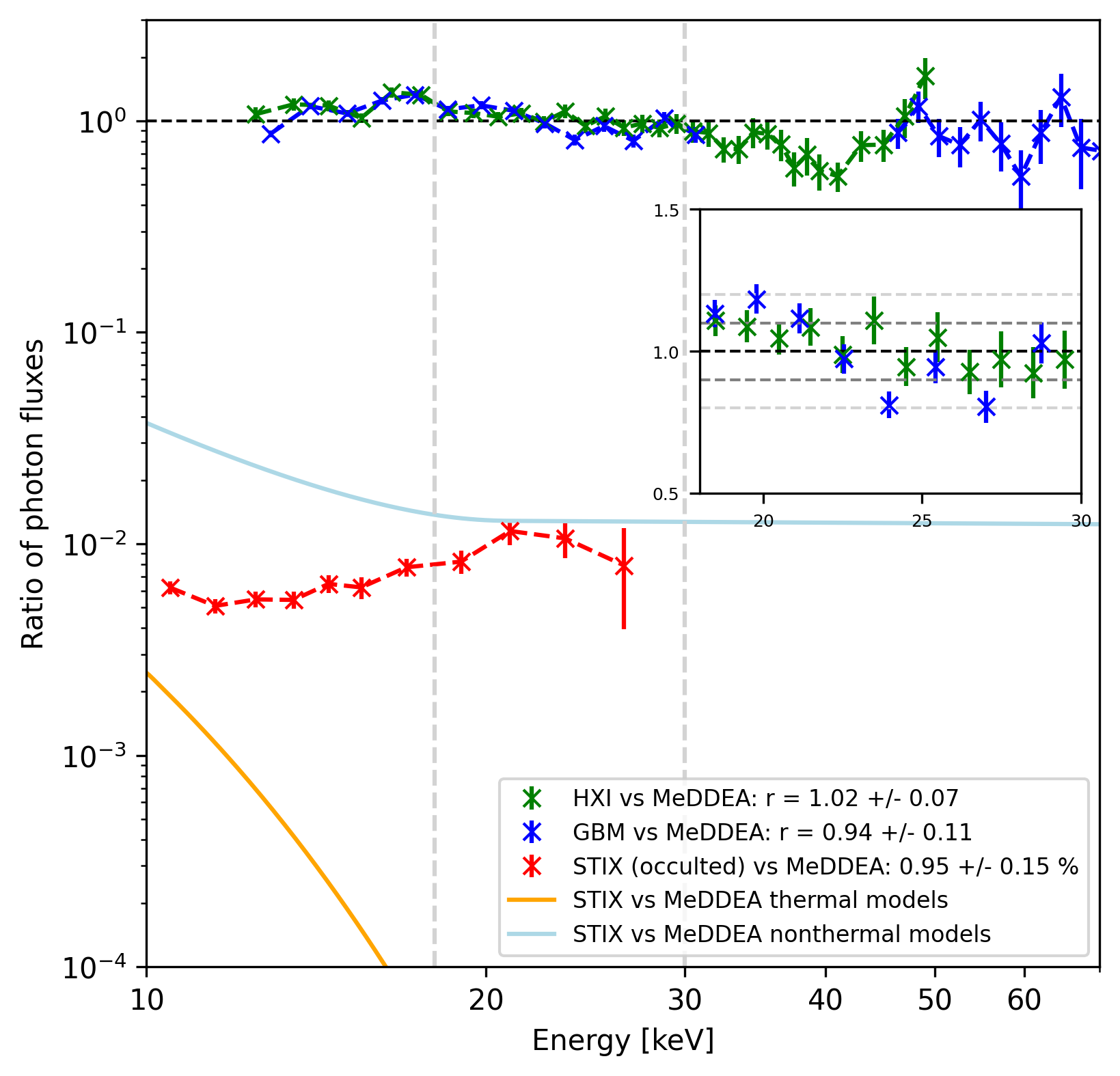}
   \caption{Ratio of the photon spectra as a function of energy. The MeDDEA spectrum is compared to the HXI (green), GBM (blue), and STIX (red) spectra. The horizontal dashed black line shows the expected 1.0~ratio between MeDDEA, HXI, and GBM. The vertical gray lines mark the region of 18--30~keV, where the three instruments align well; see text for details. The reported ratios in the legend are the mean over this energy range. In addition, the ratios of the fitted thermal and thick-target photon models from MeDDEA and STIX are shown in orange and light blue, respectively. The zoomed-in region shows the 18--30~keV energy range for the comparison between MeDDEA, HXI, and GBM. The horizontal black and gray lines mark the 0.8, 0.9, 1.0, 1.1, and 1.2 ratios.}
   \label{Fig: Ratio spectra}%
    \end{figure}

\subsection{First MeDDEA flare spectrum}\label{Sec: First MeDDEA spectrum}

Figure~\ref{Fig: Comparison spectra}(a) shows the count spectrum observed by MeDDEA of the large pixels integrated over 8~s around the nonthermal peak of the M6-class flare. We fitted a thermal component, i.e., an isothermal Maxwellian distribution model, and a nonthermal component, i.e., a thick target model \citep[e.g.,][]{Brown_1971}, to the spectrum. We did not use an albedo component in the fit, as the flare is observed at the limb, and therefore the albedo component is expected to be negligible \citep[e.g.,][]{Kontar_2006}. We subtracted a pre-flare spectrum as the background, plotted in the same panel as the flare spectrum. The turnover of the count spectrum is around 11--12~keV due to the aluminum filter (Christe et al. sub. to A\&A 2026). For medium-sized flares, this gives a wide enough energy range for reliable spectral fitting of the thermal and nonthermal components, as demonstrated in the present example.

Panels~(b) and~(c) show the observed count spectrum over the same integration time from HXI and GBM, respectively. The GBM spectra around 35~keV contain an instrumental artifact that was not removed in the standard data product, and we therefore excluded this range for spectral fitting. For HXI and GBM, the thermal spectrum is suppressed due to thicker entrance windows compared to MeDDEA, making thermal fitting more difficult, at least during the impulsive phase when the nonthermal component is strong. Considering these differences, the fitted temperatures and emission measures are different but of the same order of magnitude.
At higher energies where the nonthermal component dominates, the fits agree with similar power-law slopes. The differences between the indices are 1.9 and 2.1$\sigma$ for MeDDEA compared to HXI and GBM, respectively. In summary, the nonthermal component agrees as well as could be expected.

To quantify the agreement between the different instruments, the fitted photon fluxes are compared. Figure~\ref{Fig: Ratio spectra} shows the ratio between the MeDDEA photon spectrum and the HXI and GBM photon spectra, respectively, as a function of energy. Only the energy ranges that are fitted are shown. As the three instruments observe the flare from the same viewing angle, it is expected that the ratio is~1, denoted by the horizontal dashed line. In the energy range of 18--30~keV, marked by the two vertical dashed lines, the instruments are consistent with each other to within $\sim$10\% 
The average ratio within this energy range is 1.01 $\pm$ 0.10, computed over the ratios of both the comparison between MeDDEA and HXI and MeDDEA and GBM. The average ratio for the individual comparisons are reported in Fig.~\ref{Fig: Ratio spectra} in the legend. At energies below the peak in the count spectrum, the deviations, as visible in the plot, are larger due to calibration difficulties of HXI and GBM. 
At higher energies above 30~keV, the flare presented in this paper lacks statistics for a good cross-calibration. The presented results show that a spectral analysis with MeDDEA and its DRM works accurately within the 18--30~keV range. 

   \begin{figure}
   \centering
   \includegraphics[width=9cm]{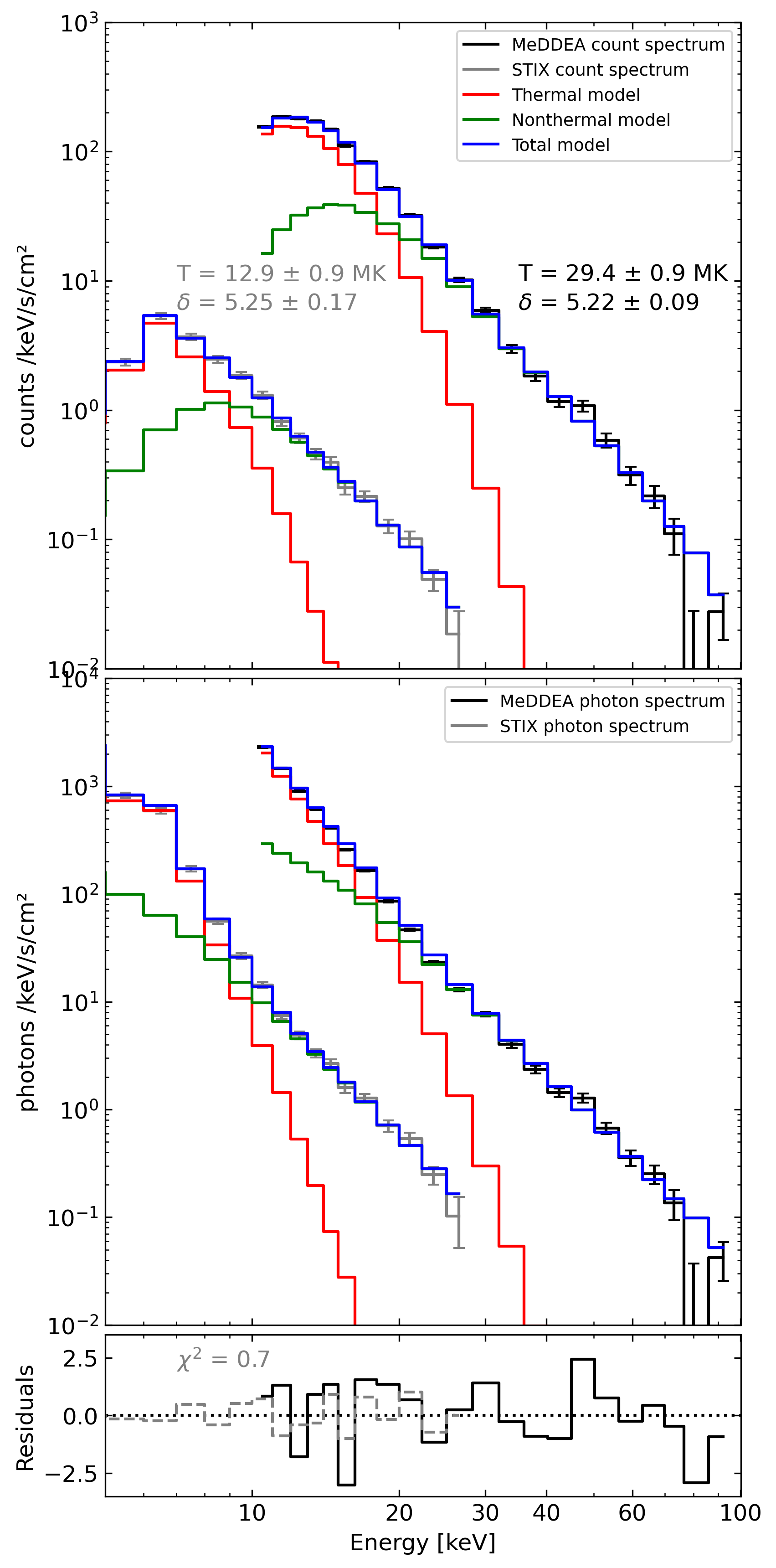}
   \caption{Comparison of the count (top) and photon (middle) spectra recorded by STIX (gray) and MeDDEA (black). A thermal (red) and a nonthermal (green) model were fitted to each spectrum. The total model is shown in blue. The temperature ($T$) and spectral index ($\delta$) are given in the top plot for the two instruments. The lowest panel shows the residuals for STIX (gray) and MeDDEA (black).}
   \label{Fig: Comparison spectra flare/CME}
    \end{figure}

\subsection{Comparing HXR flare and CME emission}

The comparison of the observed MeDDEA and STIX count spectra, as well as the comparison of the fitted photon flux, are shown in Fig.~\ref{Fig: Comparison spectra flare/CME}. While the STIX spectrum is fainter by about two orders of magnitude, the nonthermal electron spectral index is consistent with values around $\delta \approx 5.2$, the same behavior as for the event discussed in \citet{Krucker_2026}. These spectral slopes are derived using the thick-target approximation. While the thick-target approximation is generally used for flare ribbon sources, it is not straightforward that the same assumptions are valid for emission in the high corona. Using the thin-target approximation in the CME-associated spectral slope provides a harder slope in the CME component. 

The thermal fit of the flare spectrum shows a rather hot flare loop with temperatures close to 30~MK. In contrast to the previous observation, which revealed hot temperatures in the CME \citep{Krucker_2026}, the CME-associated component found here is significantly lower, i.e., around 13~MK, which is closer to temperatures reported in \citet{Hayes_2024}. 

Figure~\ref{Fig: Ratio spectra} shows the ratio between the MeDDEA and STIX photon spectra as a function of energy. In the 18--30~keV energy range, where nonthermal emission is dominant, the fraction is 0.95\% $\pm$ 0.15\%. The much lower temperature of the CME relative to the flare changes the ratio to lower values below 18 keV. To illustrate this behaviour, Fig.~\ref{Fig: Ratio spectra} shows the ratios of the thermal and thick target components separately. The increase below 18 keV in the thick target model is due to the different values of the fitted cutoff energies. As the cutoff is not constrained in the fit, the increase should not be overinterpreted. In the thick-target approximation, the electron population in the CME has the same spectral shape as in the flare, but only about 1\% of the flare-accelerated electrons' photon flux is present in the CME photon flux. For the thin-target case, the fraction of electrons in the CME increases, and it becomes energy-dependent as well. As the thin-target assumption depends on the unknown ambient density, it is difficult to obtain estimates for the thin-target case. Compared to the flare reported in \citet{Krucker_2026}, the present flare has a nonthermal component that is about five times fainter (assuming the thick-target model for both events).  

\begin{figure*}
   \centering
   \includegraphics[width=18cm]{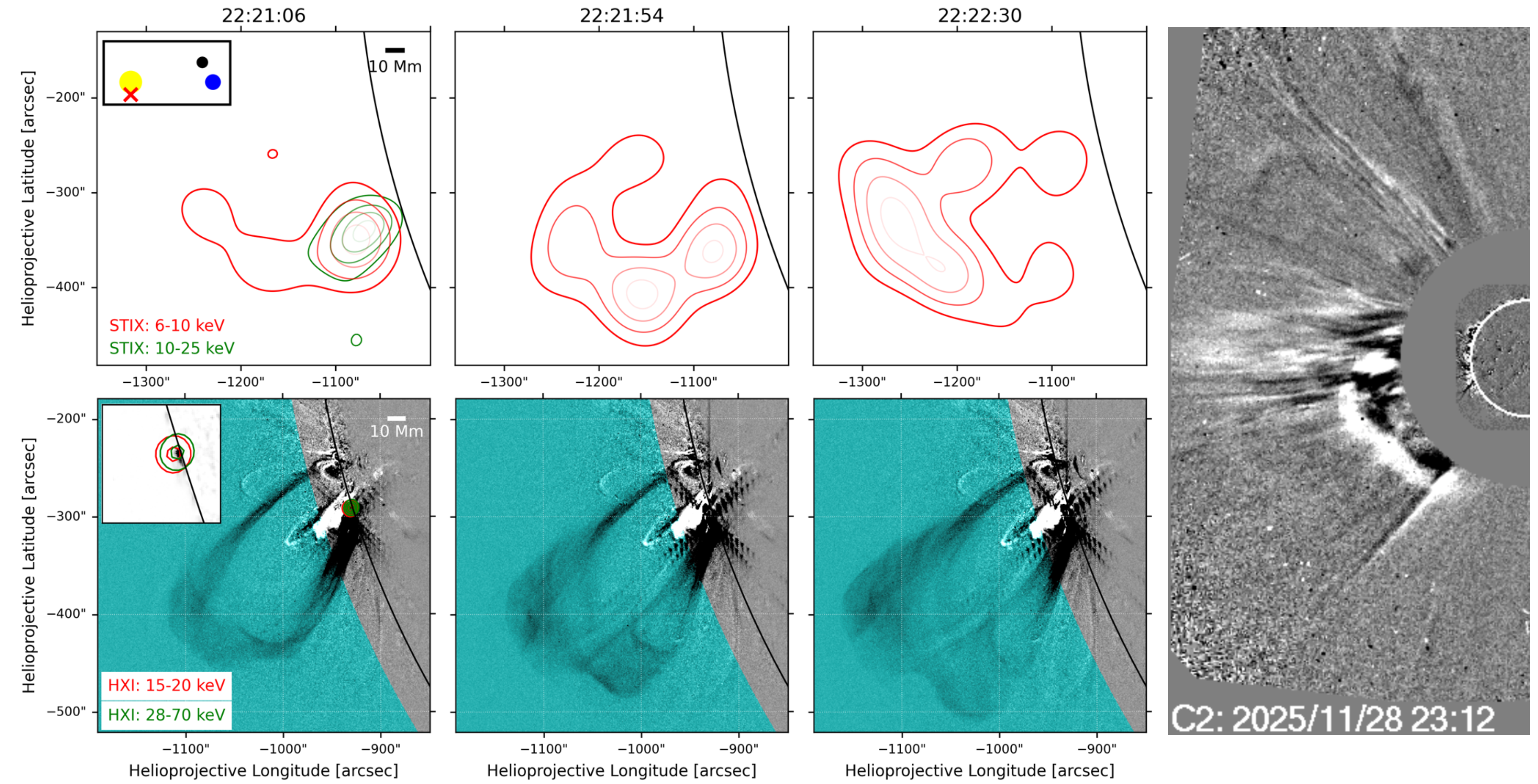}
   \caption{STIX, HXI, and AIA 131~{\AA} difference images for three different time steps representing the main HXR peak, the thermal peak, and the initial decay of the flare. The time ranges shown here are marked in Fig.~\ref{Fig: Overview lightcurves} with the red and orange bars in the STIX light curves. The top row shows the STIX reconstructions from the Solar Orbiter vantage point. The red contours correspond to 6--10~keV (thermal) and the green contour to 10--25~keV (nonthermal). In the top left panel in the left corner, the relative positions of the Sun (yellow), Solar Orbiter (black), Earth (blue), and the flare (red) are shown. In the bottom row, AIA 131~{\AA} difference images are shown. The field of view for AIA is selected to represent the same projected area as for the STIX images shown above. To guide the eye, we included a bar indicating 10 Mm in the top and bottom rows of images. The blue colored region in the AIA images approximates the region that is visible by STIX. In the first panel, HXI 15--20~keV (red) and 28--70~keV contours of the flare sources are shown. All contour levels are of 30\%, 50\%, 70\%, and 90\%. The STIX contours are shown with color shading according to the different contour levels. In the top left corner, a zoomed-in image of the HXI 50\% and 90\% contours together with an HMI difference image is shown. In the rightmost panel, the C2 difference image of LASCO is shown.}
   \label{Fig: Time evolution imaging}%
    \end{figure*}

To compare the thermal emission of the flare and the CME, we estimated the thermal energy using the following standard approximation:
\begin{equation}
    E_{th} = 3k_b T\sqrt{EM\cdot V} \qquad,
\end{equation}
where the temperature, $T$, and emission measure, $EM$, are given from spectral analysis. The volume, $V$, can be estimated using imaging results. In the first column of Fig.~\ref{Fig: Time evolution imaging}, the reconstructed images of STIX and HXI are shown for the nonthermal peak time. Volume estimates are obtained through the commonly used approach \citep[e.g.,][]{Warmuth_2013, Caspi_2014} of taking the 50\% contour area of the thermal HXR source to the power of~1.5: $V = 4/3\cdot\pi\cdot A^{3/2}$. For the CME source, we additionally used AIA 131~{\AA} images to estimate the volume of the erupting flux rope, estimating a thickness of 15{\arcsec}$\times$15{\arcsec} and a flux-rope length of 400{\arcsec} from the AIA maps. The results from the thermal energy analysis are reported in Table~\ref{Tab: Fitting results}. The error bars given are calculated using error propagation with the errors in $T$ and $EM$ from the spectral fitting, but no systematic errors from the volume estimates were considered. Furthermore, a filling factor of~one was assumed. For smaller filling factors, the energies are accordingly lower.

Considering the error bars and the spread of results using the two different methods to estimate the volume, we conclude that the thermal energy content is of the same order of magnitude for the flare and the CME-associated source, at least for a filling factor near~1. These results align with the report by \citet{Krucker_2026} and demonstrate that HXR emission from the erupting CME is an integral part of the total eruption. 

   \begin{table}
      \caption[]{Summary of the parameters as observed by MeDDEA (flare emission) and STIX (high corona emission).}
      \vspace{-1.0em}
         \label{Tab: Fitting results}
         $$
         \vspace{-0.8em}
         \begin{array}{|l|l|l|l|}
             \hline
             & & &  \\
               & \mathrm{Flare\;emission} & \mathrm{High\;corona} & \mathrm{Ratio}\\
                & & &  \\\hline
             \mathrm{T} & 29.4 \pm 0.9 & 12.9 \pm 0.9 & \\
             \mathrm{[MK]} & & & \\\hline
             
             \mathrm{EM} & 1.1 \pm 0.1 & 0.2 \pm 0.1 & \sim 0.2\\
             \mathrm{[10^{48}\;cm^{-3}]} & & & \\\hline
             
             \mathrm{V} & 3.5\times 10^{26} & (1)\;7\times 10^{28} & \sim 100\\
             \mathrm{[cm^3]} &  & (2)\;3.5\times 10^{28} & \\\hline
              
             \mathrm{E_{th}} & 2.4 \pm 0.1 & (1)\;6.1 \pm 1.6 & \sim 2\\
             \mathrm{[10^{29}\;erg]} &  & (2)\;4.3 \pm 1.1 & \\\hline

             \mathrm{n} & 5.54 \pm 0.25 & 0.16 \pm 0.04  & \\
             \mathrm{[10^{10}\;\;cm^{-3}]} &  &  & \\\hline
              
             \mathrm{\delta} & 5.22 \pm 0.09 & 5.25 \pm 0.17 & \\
            \hline
         \end{array}
         $$ 
         \tablefoot{From top to bottom: temperature, $T$; emission measure, $EM$; volume, $V$; thermal energy, $E_{th}$; density, $n$; and spectral index, $\delta$. For the volume estimate in the high corona, we used two methods: the STIX reconstructed image (1) and the AIA 131~{\AA} loop (2). The thermal energies of both methods are reported. In the last column, the rough ratios between the high corona measurement and the flare measurement are shown.}
   \end{table}

\subsection{HXR imaging in the high corona} \label{Sec: HXR imaging}

In Figure~\ref{Fig: Time evolution imaging}, the temporal evolution of the HXR source as observed by STIX is shown. For comparison, we show AIA 131~{\AA} base difference images in the row below. The color map was chosen such that dark regions represent regions of enhanced emission. The Extreme Ultraviolet Imager \citep[EUI;][]{Rochus_2020} on the Solar Orbiter was in a six-minute cadence, missing the impulsive phase, and we therefore decided not to include the data in our study.

HXI thermal and nonthermal contours are included in the first time step. It is obvious that STIX and HXI see different parts of the eruption. The flare sources seen by HXI are compact with the thermal emission lying only $\sim$2 \arcsec above the nonthermal emission. Hence, the flare loop is extremely compact with a radial height of 1.5 Mm.
On the other hand, STIX observes very large extended structures above the solar limb. As there is no detectable modulation in the finest grids for the STIX imaging system, only sub-collimators 7--10 were used for the reconstructions, in the same way as described in \citet{Hayes_2024}. 

The STIX thermal images look similar to the AIA 131~{\AA} loop structure, showing the escaping CME. To visualize this better, we calculated the occultation height of STIX, using Eq.~(1) in \citet{Krucker_2026}. Anything $\gtrsim$34{\arcsec} above the limb from the Earth's perspective is visible for Solar Orbiter, which is the blue region in the AIA 131~{\AA} images. The fields of view of the top and bottom panels in Fig.~\ref{Fig: Time evolution imaging} are adapted to each other, as indicated by the 10~Mm bar, such that structures of the same size are also displayed at the same size. 
Initially located right above the solar surface, the HXR source moves in a northeastern direction over the three time steps. The same movement is seen in AIA, where the loop is breaking up and moving in a northeastern direction. This is observed later in the LASCO C2 difference image as a CME, which is shown in the right panel of Fig.~\ref{Fig: Time evolution imaging}. By tracking the centroid position of the STIX images, we estimated the source speed over the three time steps, resulting in a velocity of around 450~km/s. This is comparable to, but at the lower edge of, the reported CME speeds of $\sim$700~km/s. The nonthermal source of STIX in the first time interval seems to be predominantly coming from the southern leg of the loop, likely due to enhanced ambient density in that region. 

\section{Discussion and conclusion}\label{sec:discussion}

In this paper, we present the first flare observed by the MeDDEA instrument on the NASA CubeSat PADRE, listed as an M6-class flare on 28 November, 2025 at around 22:20~UTC. We compared the MeDDEA spectrogram, light curves, and flare spectra with the observations of ASO-S/HXI and Fermi/GBM, two instruments that observed the Sun from the same viewing angle as MeDDEA. The time analysis has shown consistent results among the different instruments. Comparing the flare spectrum of MeDDEA integrated around the nonthermal peak with HXI and GBM showed that the three instruments are consistent within 10\% in the 18--30~keV energy range. The flux in the lower energies is more difficult to compare since HXI and GBM are less well calibrated in that range. For the cross-calibration at higher energies, a larger and harder flare is needed with sufficient statistics in all instruments at the high energy levels. In particular, a large flare jointly observed by MeDDEA and STIX is desired for a more detailed cross-calibration, as the two instruments fly the same detectors. This will be an important step for directivity measurements between MeDDEA and STIX and will be part of a dedicated future paper.

The M6-class flare was limb-occulted for STIX, such that the STIX signal is from within the erupting CME without any counts from the flare itself. Comparing the STIX signal with the MeDDEA signal in the energy range of 15--25~keV, we find that the occulted photon flux is around 1\% of the flare photon flux. Such a small fraction is very difficult to detect with current instrumentation and can only be reliably detected in limb-occulted flares. 

Assuming a thick-target model in the CME, only 1\% of the electrons actually move upwards and produce bremsstrahlung in the escaping flux rope, while for the thin-target case, the fraction can be larger. The ambient density that is required for the thick-target scenario to be appropriate can be roughly estimated from the stopping column density \citep[e.g.,][]{Emslie_1978}. Using electrons at an energy level of 30~keV as an example, we obtain a column density of $1.4\times10^{20}$~cm$^{-2}$. For the 8~s integration time used in the presented spectra, a 30~keV electron travels a distance of about $8\times10^{10}$~cm. Hence, a density above $1.7\times10^{9}$~cm$^{-3}$ is required to have a thick target scenario. This is a plausible value for a density in the high corona. The density of the thermal emission seen by STIX gives an order of magnitude density of $2\times10^{9}$~cm$^{-3}$; see Table~\ref{Tab: Fitting results}. While this is not a clear-cut result, it favors the thick-target model, at least for electrons at the low energy end of the accelerated spectrum. In a thick-target case \citep[e.g.,][]{Veronig_2004}, electrons stop in the high corona and no HXR emission in the anchor points of the erupting flux rope is produced. A few flares show hard X-ray sources from the anchor points of the erupting flux rope \citep[e.g.,][]{Stiefel_2023}. In such cases, the thin-target scenario seems to be realized with a column density low enough that electrons reach the anchor points of the erupting flux rope. In summary, it is not a priori clear which approximation to use to estimate the number of energetic electrons in high coronal sources. Furthermore, it could be the case that for some field lines within the filament the thick target is realized, while on others the column density is much lower and the thin target is the better approximation. In any case, future observations should focus on better constraining ambient densities in the flux rope.

Using the thermal parameters derived from spectral fitting and volume estimations from imaging, we were able to calculate the thermal energy in the flare and in the erupting CME. We find that the thermal energy between the two is at the same order of magnitude, even with the tendency of more energy in the CME, highlighting that HXR sources in the high corona are an integral part of solar eruptions. 

This paper presents the first scientific analysis of MeDDEA data on board the PADRE CubeSat. MeDDEA observes in an energy range ideal for flare observations of larger C-, M-, and X-class flares, covering both thermal and nonthermal flare emission. While the main science objective of PADRE with MeDDEA is to study electron beaming in flares, this paper demonstrates that MeDDEA provides high-quality flare observations that can be used for flare studies with various scientific goals. When taking advantage of the different vantage points of STIX and MeDDEA, not only do directivity studies become available, studies into the complex nature of solar eruptions through occulted observations also become possible.
 
\begin{acknowledgements}
M.Z.S. thanks everybody in the PADRE team for making the mission possible. PADRE is a NASA CubeSat mission and efforts of the PADRE team are funded under the cooperative agreement 80NSSC22M0098. Solar Orbiter is a space mission of international collaboration between ESA and NASA, operated by ESA. The STIX instrument is an international collaboration between Switzerland, Poland, France, Czech Republic, Germany, Austria, Ireland, and Italy. ASO-S mission is supported by the Strategic Priority Research Program on Space Science, the Chinese Academy of Sciences, Grant No. XDA15320000. SDO data are courtesy of NASA/SDO and the AIA. This CME catalog is generated and maintained at the CDAW Data Center by NASA and The Catholic University of America in cooperation with the Naval Research Laboratory. SOHO is a project of international cooperation between ESA and NASA.
This work was conducted using tools provided by SunPy \citep[][]{sunpy_community2020}, Astropy \citep[][]{astropy:2022}, Matplotlib \citep[][]{Hunter_2007}, and NumPy \citep[][]{harris2020array}.
M.Z.S. is supported by the Swiss National Science Foundation Grant 10006078. M.Z.S. and S.K. want to thank the Institute for Data Science at FHNW for their continued support. G.M. and CEA acknowledge the support from Île-de-France Region (grant IDF-DIM-ORIGINES-2025-2-04). L.A.H is supported by a Royal Society – Research Ireland University Research Fellowship (URF\textbackslash R1\textbackslash 241775).
We thank Marina Battaglia for her helpful comments and suggestions on the manuscript. The authors thank the anonymous referee for the helpful comments which improved the paper.
\end{acknowledgements}

\bibliographystyle{aa}
\bibliography{aa62364-26}

\end{document}